\documentstyle[preprint,aps]{revtex}
\begin{document}
\title {Potential, core-level and $d$ band shifts at transition metal surfaces}
\author{M.V. Ganduglia-Pirovano\footnote{Present address:
Center for Atomic-Scale Materials Physics and
Physics Department, Technical University of Denmark,
DK-2800 Lyngby, Denmark}}
\address{Max Planck Institut f\"ur Physik komplexer Systeme, Au\ss enstelle
Stuttgart, Heisenbergstrasse 1, D-70569 Stuttgart, Germany\\and\\
Fritz-Haber-Institut der Max-Planck-Gesellschaft, Faradayweg
4-6, D-14$\,$195 Berlin-Dahlem, Germany}
\author{V. Natoli and M. H. Cohen}
\address{Corporate Research Science Laboratories, Exxon Research and 
Engineering Company, Annandale, New Jersey 08801}
\author{J. Kudrnovsk\'{y}}
\address{
Institute of Physics, Academy of Sciences of the Czech Republic,
CZ-180$\,$40 Prague 8, Czech Republic\\and\\
Institute of Technical Electrochemistry, Technical University, A-1060
Vienna,
Austria}
\author{I. Turek}                   
\address{Institute of Physics of Materials, Academy of Sciences of the Czech 
Republic, CZ-616$\,$62 Brno, Czech Republic}

\date{\today}
\maketitle

\begin{abstract}
We have extended the validity of the correlation between
the surface $3d$ core-level
shift (SCLS) and the surface $d$ band shift (SDBS) to the entire $4d$
transition metal series and to the neighboring elements Sr and Ag via
accurate first-principles calculations. We find that the correlation is
quasilinear and robust  with respect to the differencies both between initial
and final-state calculations of the SCLS's and two distinct
measures of the SDBS's. We show that despite the complex spatial dependence
of the surface potential shift (SPS) and the location of the $3d$ and $4d$ 
orbitals in different regions of space, the correlation exists because
the sampling of the SPS by the $3d$ and $4d$ orbitals remains similar.
We show further that the sign change of the SCLS's  across the transition
series does indeed arise from the $d$ band-narrowing mechanism previously
proposed. However, while in the heavier transition metals the predicted
increase of $d$ electrons in the surface layer relative to the bulk arises
primarily from transfers from $s$ and $p$ states to $d$ states within the
surface layer, in the lighter transition metals the predicted decrease of
surface $d$ electrons arises primarily from flow out into the vacuum.

\end{abstract}

\narrowtext
\section{INTRODUCTION}

The surface electronic structure of a metal plays a central role in the
surface chemistry that takes place upon it. That is, the surface electronic
structure of a metal is central to its chemical reactivity.
It has long been understood that in the transition metals in particular,
the $d$ electrons dominate their chemistry.
More recently, computational studies of chemisorption and dissociation on      
various metals and alloys have clarified how they do so 
\cite{hammerscheffler,norskovsci,norskovnat,morikawa,steffen1,steffen2}.
Thus, one can state with confidence that the  surface chemical
reactivity of a transition metal depends strongly on the response of
its surface $d$ electrons to the external perturbations imposed by an
atom or molecule with which it interacts.
A generic measure of its reactivity should then be provided by their response
to a generic perturbation.

One very simple generic perturbation is provided by the difference in 
environment between surface and bulk. Thus, a shift in some suitable feature
of the $d$ band structure between surface and bulk could provide a suitable
generic measure of reactivity. The structure of the occupied portion of the
$d$ band could be studied experimentally by photoelectron spectroscopy, and
the shift in the peaks of the spectrum from surface to bulk could be detected.
This measure of the surface $d$ band shift (SDBS) is somewhat ambiguous because
of broadening and the role of matrix elements. A sharply defined measure
is the shift in the center of gravity of the $d$ band, $B_d$, or

\begin{equation}\label{eq0}
{\rm SDBS}\approx \Delta B_d= -(B_d^{surf}-B_d^{bulk})\quad .
\end{equation}

\noindent Another sharply
defined measure would be the shift in the relevant matrix
element of the Kohn-Sham Hamiltonian; how to model this
is discussed in Sec. III.

A recent development which is very interesting is that the 
differences between the 
core-electron binding energies of  surface atoms of supported
metal monolayers on
transition-metal substrates  and of surface atoms of the clean
elemental crystal surface (the
latter consisting of the  element forming the adlayer)
 are strongly correlated with the corresponding
shift in the center of gravity of the surface $d$ band   
\cite{veronica,weinert}.  It has been observed 
\cite{goodman1} that the core-level shifts tend to correlate with the adlayer's chemical
reactivity and that this could be understood through the correlation with
the surface $d$ band shift  \cite{veronica,morikawa}. 

The SCLS's and the SDBS's arise from the surface potential shift experienced
by the core states and the $d$ states, respectively. Both
represent different spatial samplings of the same potential shift; thus the
existence of the observed correlation is not surprising. The purpose
of this paper is to provide a quantitative test of the correlation and a     
deeper and more detailed understanding of its origin.
In Sec. II, we report the results of SCLS calculations for the $4d$
transition metals and for Sr and Ag. We also 
demonstrate 
an excellent
quasilinear correlation between the computed SCLS values and those of
the SDBS's. Indeed, the SCLS values nearly equal  the SDBS values in the   
initial-state approximation. To demonstrate that this correlation
is robust, holding for other measures of the SPS, we extract values of 
that shift for the $xy$ and $x^2-y^2$ subbands at the center of the
surface Brillouin zone by fitting the computed bands to a simple model
in Sec. III. The correlation is found to persist.

The two most prominent features of the results in Sec. II and III are
({\it i}) the quasilinear correlation between the SCLS's and the SDBS's
and ({\it ii}) the sign change which occurs across the $4d$ series.
Understanding ({\it i}) requires less depth of analysis than understanding
({\it ii}). Accordingly, we present in Sec. IV plots of the $3d$ core-orbital
density and the $4d$ valence-orbital density overlaid upon the SPS
which make clear that the sampling argument referred above is valid.
On the other hand, to understand ({\it ii}) fully requires probing more
deeply into the origin of the SPS than is done in the
usual surface $d$-band-narrowing argument.
In Sec. V we provide 
such an analysis of the
origin of the SPS
via a decomposition of the SPS into its component parts together with a
decomposition of the electron-density changes $\Delta \rho(r)$
responsible for it. In the process, we confirm the essential correctness of
the band-narrowing argument. The resulting improved understanding of the SPS
and $\Delta \rho(r)$ is our most significant contribution.
We discuss our results briefly in Sec. VI.

\section{CORRELATION BETWEEN SCLS's and SDBS}

We have considered the fcc(001) surfaces of the $4d$
transition metals and of Sr and Ag and have calculated the SCLS's,
$\Delta_c$, for the $3d$ levels.
All calculations have been done for the fcc(001)
structure to facilitate intercomparison and the establishment of trends,   
irrespective of whether fcc is the equilibrium structure.       

Using Slater's transition-state concept \cite{ts}
to evaluate total-energy differences, $\Delta_c$ can be estimated from

\begin{eqnarray}
\Delta_c & \approx &
-[\epsilon_{c}^{\rm surf}(n_{c}=1.5)
-\epsilon_{c}^{\rm bulk}(n_{c}=1.5)]\quad , \label{eq1}
\end{eqnarray}
\noindent where 
$\epsilon_{c}^{\rm surf}$ and $\epsilon_{c}^{\rm bulk}$
denote the
Kohn-Sham eigenvalues of a particular core state of a surface or bulk atom,
and $n_c$ is the occupation number of the core-orbital.
In the initial-state approximation, the SCLS's
are given by Eq. (\ref{eq1}) with $n_c=2$.

The electron density, core-electron eigenvalues, and densities of states are
calculated by means of an efficient surface-Green's-function technique based on the tight-binding, all-electron
linear muffin-tin orbital
theory within the local density approximation (LDA)
of the exchange-correlation functional in the Ceperley and Alder form
\cite{ceperley} as parametrized by Perdew and Zunger \cite{perdew}.
The details of the method have been described
elsewhere \cite{josef1,josef2}.
The potentials are calculated selfconsistently
within the atomic-sphere
approximation (ASA) in an intermediate region
consisting of the
surface layer, three substrate layers, and two layers of
empty spheres simulating the vacuum-sample interface. This intermediate
region is coupled to the semi-infinite vacuum on one side
and to the semi-infinite crystal on the other, with frozen potentials.
Calculations are performed for 
sphere radii chosen  so as to minimize the total energy of the bulk in
the fcc structure.
No relaxation of the topmost layers has been assumed; all
interatomic distances in the surface layer, as well as between surface and
substrate layers are assumed to be the same and equal to those in the substrate.
To describe the transition-state (see Eq. \ref{eq1}), selfconsistent 
electronic structure calculations are performed under the constraint of 
charge neutrality. The single impurity problem of the localized core-hole with
half an electron missing in the $3d$ shell can be easily treated within
the surface Green's function formalism. The charge neutrality constraint
within this scheme results in a nearly fully relaxed final state, i.e.,
that the valence charge of the excited single impurity nearly equals 
(Z+0.5)$e$ (Z atomic number).
In Table \ref{table1} we compare three of our calculated shifts with 
independent first-principles calculations, which include final-state
effects to some approximation, and with available
experimental data. A number of other calculations for the most close-packed
surfaces \cite{alden} have achieved a similar degree of agreement
between experiment and theory using similar but not identical computational methods. 
 
We have also calculated the LMTO potential parameters $C_d$
\cite{lmto}, which correlate
closely with the center of gravity of the $d$ band, $B_d$, for both bulk
and surface and have constructed the SDBS's from
them according to Eq. (\ref{eq0})
$(\Delta B_d\approx \Delta C_d)$.
The calculated initial-state and final-state core-level shifts for the
fcc(001) surfaces of the $4d$ transition metals plotted vs. these calculated
SDBS's are presented in Fig. \ref{fig1}. Note the near linearity of the         
correlation between the initial SCLS's and the SDBS's.
The smaller screening contributions to
the shifts, which are related to the
downward shifts of the core-levels when deoccupied
leave the overall
linear correlation largely unaffected.
                 
\section{CORRELATION OF THE SCLS's WITH SURFACE POTENTIAL PERTURBATIONS}

We have somewhat arbitrarily selected the SDBS as defined by Eq. (\ref{eq0})
as the generic measure of the response of the $d$ electrons to a generic
perturbation and have shown that it has a beautiful quasilinear correlation
with the SCLS's. However, the $d$ band structure is complex, with changes
in width, shape and centroid position at the surfaces. 
We now examine the relationship between the SCLS's and the values of a quite different measure
of the SDBS to test whether the above correlation is relevant despite the complexity of the 
$d$ band.

In earlier papers we have shown how to extract and display the individual   
physical effects of the presence of the surface on the electronic structure
by examining the ${\bf k_{||}}$-, symmetry-, and layer-resolved density of
states (DOS) at ${\bf k_{||}=0}$ for both clean surfaces \cite{veronica1} and 
overlayers \cite{veronica2}. For the fcc(001) surface, the 
$xy$ and $x^2-y^2$ subbands do not couple with any other low-lying bands.
The bulk $xy$ and $x^2-y^2$ contributions to the DOS are accurately fitted
by a simple cosine band and are thus 
represented by the band structure of an infinite 
one-dimensional chain of sites having
only nearest-neighbor coupling, $t$, of a non-degenerate level.
For the surface layer and substrate layers underneath, the $xy$ and $x^2-y^2$
subbands DOS at ${\bf k_{||}=0}$, are accurately represented by that of a 
semi-infinite chain perturbed by potential shifts $V_1$ and $V_2$ at the    
terminal and penultimate sites.
The quantity $V_1$ can be interpreted as a model of the
surface shift of the diagonal matrix element of the true Hamiltonian in a
generalized Wannier representation. We take its negative as an alternative
measure of the SDBS which reflects in quantitative  detail the shift in the
surface $d$ band structure caused by the surface potential shift.
The band widths $4t$ were fixed by fitting to the bulk subbands and the values
of $V_1$ and $V_2$ were fixed by fitting the zeroes in the DOS in the third
layer. The resulting fits for the full energy and layer
dependence were in general extraordinary as shown in Fig. \ref{fig3} for the
Rh $xy$ subband, for which the fitted and the calculated DOS are nearly
undistinguishable.
In Ref. \cite{veronica1}, the DOS was fitted with $V_1$ only.
The fits with two parameters are  slightly better, but the 
$V_1$ values are robust, i.e. they change little between the one and
two parameter fits.
The calculated initial-state and final-state core-level shifts
for the fcc(001) surface of the $4d$ transition metals plotted vs. the
fitted surface potential shifts, $V_1$,
for the $x^2-y^2$, and $xy$ subbands, 
are presented in Figs. 2a and 2b, respectively.
Once again we find a quasilinear correlation between the
initial-state shifts and an independent measure of the SDBS's which is relatively little
affected by final-state contributions to the
core-level shifts, demonstrating that the correlation
is indeed robust.

\section{THE SURFACE POTENTIAL SHIFT AND THE ORIGIN OF THE CORRELATION}

It is very interesting that the numerical values of the SCLS's are not
merely correlated with, but they are nearly equal to those of the potential
shifts $-V_1(xy)$, $-V_1(x^2-y^2)$, and $\Delta C_d$, and particularly so for
the initial-state contributions to the shifts and $\Delta C_d$. To
understand better this nearly equality we have examined the spherically 
symmetric part of the
surface potential shift $\Delta V(r)$,

\begin{equation}
{\rm SPS}=\Delta V(r)=V^{surf}(r)-V^{bulk}(r)
\end{equation}

\noindent together with $r^2 |R_{3d}(r)|^2$ and 
$r^2 |R_{4d}(r)|^2$, where $R_{3d}(r)$  and $R_{4d}(r)$         
are the radial solutions of the Schr\"odinger equation within the
corresponding atomic sphere in the bulk.
To first order,

\begin{eqnarray}
\Delta_{3d}^{initial}
 & \approx & -\int dr\, \Delta V(r)\, r^2 |R_{3d}(r)|^2\\\label{eq5}
\Delta C_d & \approx & -\int dr \, \Delta V(r)\, 
r^2 |R_{4d}(r)|^2\quad ,\label{eq6} 
\end{eqnarray}

\noindent hold, with appropriate normalization. Eq. (4) is expected
to be more accurate than Eq. (5) since the $3d$ core-level is much 
more tightly bound and the LMTO-orbital by itself cannot represent the full
complexity of the $4d$ band.

We show our results in Fig. 4a for 
Y, representative of the lighter $4d$ transition
metals with a positive shift, in Fig. 4b,
for Mo, representative of the mid series elements with small
shifts, and in Fig. 4c for
Pd, representative of the heavier $4d$
transition metals with a nearly filled $d$ band and a negative shift,
We compare the computed values for the initial-state contributions to 
the shifts and $\Delta C_d$ with those estimated via Eqs. (4) and     
(5) in Table \ref{tableII}.
We see that the agreement is better for the initial-state 
SCLS than for the $\Delta C_d$'s, as
anticipated, but Figs. 4 and Table \ref{tableII} clearly imply that the
sampling argument for explaining the correlation is correct. It could
not be anticipated in advance of these plots, however, that the sampling
argument must be correct, considering the complexity of the spatial dependence
of $\Delta V(r)$ and the fact that the $3d$ and $4d$ orbitals are localized
in quite different regions of space. However, the spread of the $4d$ orbitals
effectively averages over the spatial fluctuations of $\Delta V(r)$.

It should be noted that for radii $r$ larger than half the nearest neighbor separation,
indicated by a vertical line in both Figs. 4 and 5, the computed quantities become 
unreliable because of the use of the ASA. The resulting uncertainties do
not weaken any of the conclusions drawn above from Figs. 4 and below from Figs. 5.

\section{ORIGIN OF THE SURFACE POTENTIAL SHIFT}

Figs. 1, 2a and 2b show the well-known change of sign that was observed
to occur in the middle of the $5d$ transition series \cite{veen}
and predicted for the most close-packed surface of the 
observed crystal structure of the $4d$ series \cite{alden}.
This fact has usually been 
qualitatively explained by using the decrease in width of the $d$
band at the surface compared to the bulk and assuming an approximate
conservation of $d$ charge
in each layer. Thus, the self-consistent potential at the surface changes 
so as to mantain the $d$ band filling approximately constant, and consequently
the surface
$d$ band shifts relative to the 
bulk band. This perturbing potential acts on the core
electrons as well and is repulsive for late and attractive for early
transition metals. 
These arguments apply as well to the nominally empty or filled
$d$ bands of Sr and Ag \cite{citrin}.
For Ag, the $sp$ states below and above the Fermi level, $E_{\rm F}$,
are hybridized with the $d$ states below $E_{\rm F}$. The $d$ band narrows
at the surface, as above. Were the $d$ band center to remain unchanged,
the $sp-d$ hybridization of the empty states above $E_{\rm F}$ would decrease,
and the net $d$-character $f_d$ of the occupied states would increase.
Here $f_d$ is defined as $f_d=\int^{E_{\rm F}} dE\, n_d(E)$, where $n_d(E)$ is 
the density of states per atom projected on to the LMTO $d$ basis functions,
either for a bulk or a surface site. A repulsive potential shift moves the
$d$ band center upward and conserves the $d$ character. For Sr, the $sp$
states are hybridized with states from the empty $d$ 
band above $E_{\rm F}$. When the empty $d$ band narrows, $f_d$ would decrease if the
 center of the $d$ band were to remain fixed. An attractive potential shift
moves the $d$ band center downward and approximately conserves $f_d$.

It is important to recognize that in this argument the potential shifts are 
assumed to occur primarily to conserve $d$ character and not strictly to
preserve electrical neutrality. For the heavier transition metals, 
it is known \cite{vero} that the outflow
of electrons from the surface layer to the vacuum,
which generates the surface dipole layer, 
indeed originates predominantly from states of $s$ and $p$ character.
Despite the qualitative consistency of the $d$ band-narrowing argument,
we regard it as incomplete and now present
a complementary but deeper analysis.
The new analysis yields a considerably more detailed understanding of the
surface potential shift and in the process confirms the essential features of
the band-narrowing argument.

The Kohn-Sham potential $V_{\rm TOT}$ is conveniently decomposed into three components 
in our computational methods, the Coulomb term $V_c$, which is the Coulomb
potential within a muffin-tin sphere arising from all charge within that sphere, the
Madelung term $V_{\rm M}$, which is the Coulomb potential arising from all charge external
to that sphere and which is constant within the sphere, and the exchange-correlation
term $V_{xc}$ which is evaluated in the LDA. Thus, the change in the total
potential is

\begin{equation}
\Delta {V_{\rm TOT}}= \Delta {V_{\rm M}}+\Delta {V_c}+\Delta {V_{xc}}
\end{equation}

\noindent where the $\Delta$ indicates the difference between surface and bulk 
quantities. The $\Delta {V_{\rm M}}$ are constants and are listed in Table III for
Y, Mo and Pd. The spherical averages of $\Delta {V_{\rm M}}+\Delta{V_c}$,
 $\Delta {V_{xc}}$,
and $\Delta {V_{\rm{TOT}}}$ are shown in the upper panel of Figs. 5a-c.
 One sees inmediately that $\Delta
V_{xc}$ is of significance only in the outer region of the atomic
sphere, where it is
repulsive. This occurs because  
to lowest order the derivative

\begin{equation}\label{eq9}
\Delta {V_{xc}}(r)=\left.
 \frac{\partial {V_{xc}}}{\partial \rho}\right|_r \Delta \rho(r)
\end{equation}

\noindent 
diverges as the density goes to zero, and the density is lowest in the
outer regions of  
the sphere. However, comparing $\Delta {V_{xc}}(r)$
in Figs. 5 with the $3d$ and $4d$ orbital densities in Figs. 4, one sees that
$\Delta {V_{xc}}(r)$ makes little contribution to the SCLS's and makes a 
comparable negative contribution to the SDBS's in all cases. 
$\Delta  V_{\rm M}$ is positive
 and $\Delta  V_c$ is negative in all three cases.
Moreover,
 both $\Delta {V_{\rm M}}$ and $\Delta {V_c}$ vary monotonically in the
series Y, Mo, Pd as illustrated by the values of $\Delta{V_{\rm M}}(r)$ and of 
$\Delta{V_{xc}}(r)$ 
at $r=0$ and $r=r_s$ ($r_s$ is the atomic sphere radii within the
ASA) listed in Table III and by the curves in Figs. 5.
It is only the sum of the Coulomb and
 Madelung potential shifts which establishes
the trend in the SPS's, the SCLS's, and the SDBS's.

This behavior of the potential shifts is associated with a corresponding selfconsistent 
behavior of the spherically averaged electron density $\rho(r)$. The shifts 
$\Delta \rho(r)$ are plotted in the lower panels of Figs. 5 for Y, Mo and Pd,
respectively, together with their decomposition, $\Delta \rho_l(r)$, into contributions
from states of given angular momentun, $l=s$, $p$, $d$.
One sees that the $s$ and $p$ contributions to $\Delta \rho(r)$ are everywhere
negative. Table IV lists the electron number shifts 

\begin{equation}
\Delta Q_l= 8\pi \int_0^{r_s} r^2\, \Delta \rho_l(r)\, dr
\end{equation}

\noindent and the total shift per sphere 

\begin{equation}
\Delta Q=\sum_l \Delta Q_l
\end{equation}

\noindent for the surface layer together with $\Delta Q$ for the first vacuum layer. One
sees first that the $\Delta Q$'s are essentially equal and of the opposite 
sign for the
vacuum and surface layers. That is, the net electron transfer occurs primarily between
the surface and the vacuum, with little transfer occurring between the surface
and the interior. Next, one sees that for Pd

\begin{equation}
|\Delta Q_s+\Delta Q_p|>|\Delta Q|
\end{equation}

\noindent in the surface layer. Thus, for Pd, $s$ and $p$ electrons 
both flow out
of the surface layer and are transferred into the surface
$d$ bands, giving rise to the
positive $\Delta Q_d$. For Mo, there is less than one third as much internal
transfer from $s$ and $p$ to $d$, leading to a 
small $\Delta Q_d$. Finally, for Y, $d$ electrons flow out of the
surface into the vacuum, giving rise to a  negative
$\Delta Q_d$. 

This trend in $\Delta Q_d$ is precisely what is expected
from the $d$ band-narrowing argument. 
Because for Pd the band centroid is below $E_{\rm F}$, the narrowing of the surface
$d$ band initiates a transfer of $s$ and $p$ electrons into $d$ electrons 
states at the surface which tends to
increase the Coulomb repulsion $V_c$
and  shift the $d$ band centroid upward thus
moderating but not eliminating the transfer. 
For Mo, with $E_{\rm F}$ near the centroid, the effect is much smaller. For Y, with
the centroid above $E_{\rm F}$, the narrowing causes a decrease of the $d$ electrons,
reducing $V_c$ and shifting the centroid downwards. 
One cannot at this point distinguish between a transfer of $d$ to $s$ and $p$
electrons which then flow outward and a direct outflow of $d$ electrons; the 
effect is the same. However, the large value of $r^2 R_{4d}^2(r)$ in the outer region of the
atomic
sphere evident in Figs. 4a for Y suggests at least some direct $d$ outflow.

Examination of  $\Delta \rho_d(r)$ in the density-shift plots in  the lower panels
of Figs. 5 reveals more
interesting results.
One sees clearly that both transfer from $sp$ to $d$ states
into the inner region of the atomic spheres
and $d$ flow out of the outer region into the vacuum occur in all three cases. In Pd
transfer within the interior dominates, in Mo the two effects nearly balance, and in Y
the outflow dominates.

\section{SUMMARY AND DISCUSSION}

We have established in Secs. II and III a robust quasilinear correlation between the
surface core-level shifts and two different measures of the surface potential shift
of the entire $4d$ transition series plus Sr and Ag. We then demonstrated in Sec. IV for
the representative elements Y, Mo and Pd that the initial-state contributions to the
SCLS's are accurately given by the
average of the SPS over the $3d$ LMTO-orbital and that the SDBS is given approximately
by the corresponding average over the $4d$ orbital. The latter, being quite broad,
averages over the spatial fluctuations in the SPS, and
so, in effect, samples a potential shift little different from that sampled
by the $3d$ orbital. This provides a detailed explanation
of  the correlation between the SCLS's and the SDBS's
and does not assume 
an  approximate spatial constancy of $\Delta {V}(r)$ throughout the region
sampled by both orbitals. Finally, we show in Sec. V that the sign change in the
shifts across the $4d$ series is indeed correctly given by the standard band-narrowing
argument, but that the situation is considerably more complex than envisaged in the
argument in its original form. In the heavier elements, the $s$ and $p$ character of
 the surface electron density is reduced both by flow into the vacuum and
local transfer into surface $d$ states. That is, transfer from $s$ and $p$ states provides
the increase in the $d$ electrons predicted by the band-narrowing argument. In
the lighter elements, it is the outflow of $d$ electrons into the vacuum which
provides the decrease in $d$ electrons predicted by the band-narrowing argument.
In the middle of the series, there is little net change in the number of $d$ electrons
because the 
transfer from $s$ and $p$ electrons into $d$ states at the surface
is balanced by the outflow into the vacuum 
It is this
systematic variation in the $d$ electron density shift and the shift in the
total number of $d$ electrons
 which is responsible for the sign change in the SCLS's and the SDBS's through
 its contribution to the shift in the total Hartree potential
$\Delta  V_{\rm M}+\Delta {V_c}$.

In the introduction, we pointed out that the SDBS could be regarded as a response of 
the $d$ band to a surface perturbation, $\Delta {V_{\rm TOT}}(r)$, and as such could
be used as a measure of surface chemical reactivity. However, it is a very crude
measure, indicating intensity of response but giving no indication of the
spatial variation or the
geometry of the response. Recently, some progress has been made in understanding the
geometry of response as well as its intensity through the introduction of 
the concepts of chemical reactivity theory into the discussion of the
surface chemistry of metals \cite{vero,morrel1,morrel2,steffen3}. The success of the
present study of trends in the SCLS's and SDBS's suggests that studies of trends in
these more sophisticated measures of chemical reactivity could be both feasible
and fruitful.

M. V. Ganduglia-Pirovano thanks P. Fulde
for his hospitality at the Max Planck Institute in Stuttgart,
where part of this work has been carried out. 
One of us (J.K.) acknowledges the financial support from the Grant
Agency of the Academy of Sciences of the Czech Republic (Project No. 110 437)
and the Austrian Science Foundation (Project P10231).

\begin{table}
\begin{tabular}{c l l }
&\multicolumn{2}{c }{SCLS (eV)}\\
&Calculated&Experiment\\\cline{2-3}
Rh(001)     &$-0.83^a$,$-0.62^b$   &$-0.62^c$ \\
Pd(001)     &$-0.34^a$,$-0.33^b$    &$-0.44^d$ \\
Ag(001)     &$-0.11^a$,$-0.10^b$   &$-0.0\pm 0.1^e$ [fcc(111)]
\end{tabular}
\caption{Compa\-rison be\-tween in\-de\-pen\-dent first prin\-ci\-ples
calculations of surface core-level shifts 
with available experimental data.}
\label{table1}
$^a$Present work\\
$^b$Methfessel {\it et al.} (Ref. \protect{\cite{dieter2}})\\
$^c$Borg {\it et al.} (Ref. \protect{\cite{borg}})\\
$^d$Nyholm {\it et al.} (Ref. \protect{\cite{nyholm92}})\\
$^e$Andersen {\it et al.} (Ref. \protect{\cite{dieter3}})\\
\end{table}

\begin{table}
\begin{tabular}{c  c c |c c}
&\multicolumn{2}{c|}{$1^{st}$ order perturbation theory}
&\multicolumn{2}{c }{Computations}\\
& SDBS& $\Delta_{3d}^{initial}$&
 SDBS& $\Delta_{3d}^{initial}$\\\cline{2-5}
Pd & $-0.031$& $-0.042$ & $-0.042$& $-0.046$\\
Mo & $+0.010$& $+0.013$ & $+0.011$& $+0.018$\\
Y & $+0.021$& $+0.044$ & $+0.038$& $+0.038$
\end{tabular}
\caption{Comparison of the first order estimates for the 
$3d$ core-level and $d$ band shifts
 with computed initial-state shifts and $d$ band shifts,
 in Rydbergs.}\label{tableII}
\end{table}

\begin{table}
\begin{tabular}{c c c c | c c c}
&\multicolumn{3}{c|}{$r=0$}
&\multicolumn{3}{c }{$r=r_s$}\\
& $\Delta {V_{\rm M}}$&
 $\Delta {V_c}$&
 $\Delta {V_{\rm M}}+\Delta {V_c}$&
 $\Delta {V_{\rm M}}$&
 $\Delta {V_c}$&
 $\Delta {V_{\rm M}}+\Delta {V_c}$\\
\cline{2-7}
Pd & $+0.18$& $-0.16$ & $+0.02$& 
 $+0.18$& $-0.14$ & $+0.04$\\
Mo & $+0.26$& $-0.26$ & $+0.00$& 
 $+0.26$& $-0.28$ & $-0.02$\\
Y & $+0.16$& $-0.18$ & $-0.02$&
 $+0.16$& $-0.21$ & $-0.05$\\
\end{tabular}
\caption{Potential shifts in Rydbergs at the center and surface of the
muffin-tin sphere.}
\end{table}

\begin{table}
\begin{tabular}{c c c c c | c }
&\multicolumn{4}{c|}{Surface}
&Vacuum\\
&\multicolumn{3}{c}{$\Delta Q_l$}&
$\Delta Q$&$\Delta Q$\\
& $s$&$p$&$d$&\\
\cline{2-6}
Pd & $-0.090$& $-0.227$ & $+0.077$& $-0.240$& $+0.248$\\            
Mo & $-0.075$& $-0.330$ & $+0.022$& $-0.382$& $+0.385$\\            
Y & $-0.035$&$-0.170$ & $-0.117$& $-0.322$& $+0.300$\\            
\end{tabular}
\caption{Shifts in the number of electrons per atom at the surface and vacuum layers.}
\end{table}

\begin{figure}[h]
\caption{
The calculated initial-state and final-state core-level shifts and the 
the shift of the $d$ band center between the bulk and
surface $d$ bands.
}\label{fig1}
\end{figure}
\vspace*{6.0cm}
\begin{picture}(3,10)
\put(.8,.0){\includegraphics{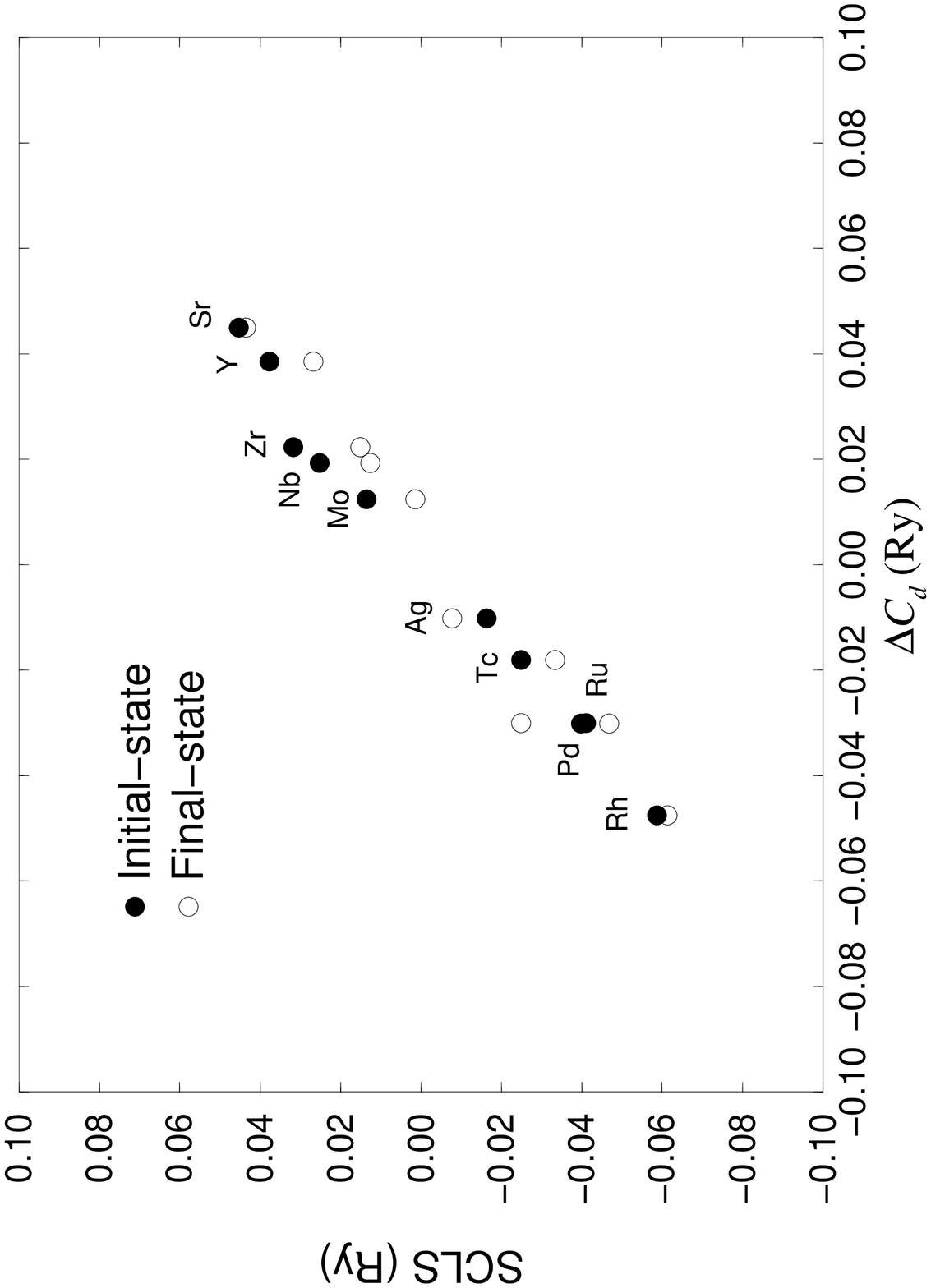}}
\end{picture}

\newpage
\vspace*{9.5cm}
\begin{picture}(3,10)
\put(.8,.0){\includegraphics{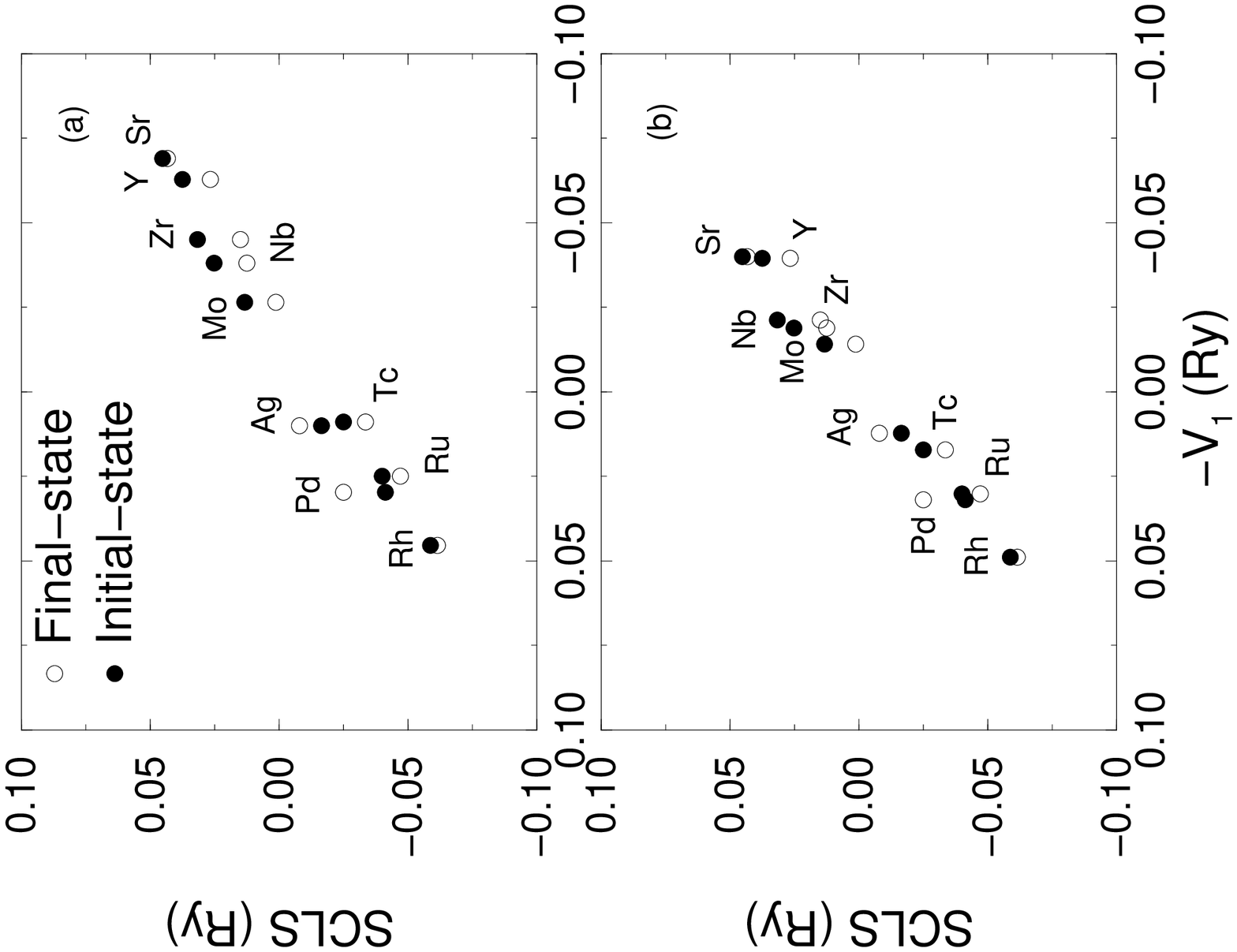}}
\end{picture}
\vspace*{-2.5cm}
\begin{figure}[h]
\caption{
The calculated initial-, and final-state core-level shifts and the fitted     
surface potential shifts, $V_1$, for the (a) $x^2-y^2$ and (b) $xy$ subbands.
}\label{fig2}
\end{figure}

\vspace*{4.5cm}
\begin{picture}(3,10)
\put(.0,.0){\includegraphics{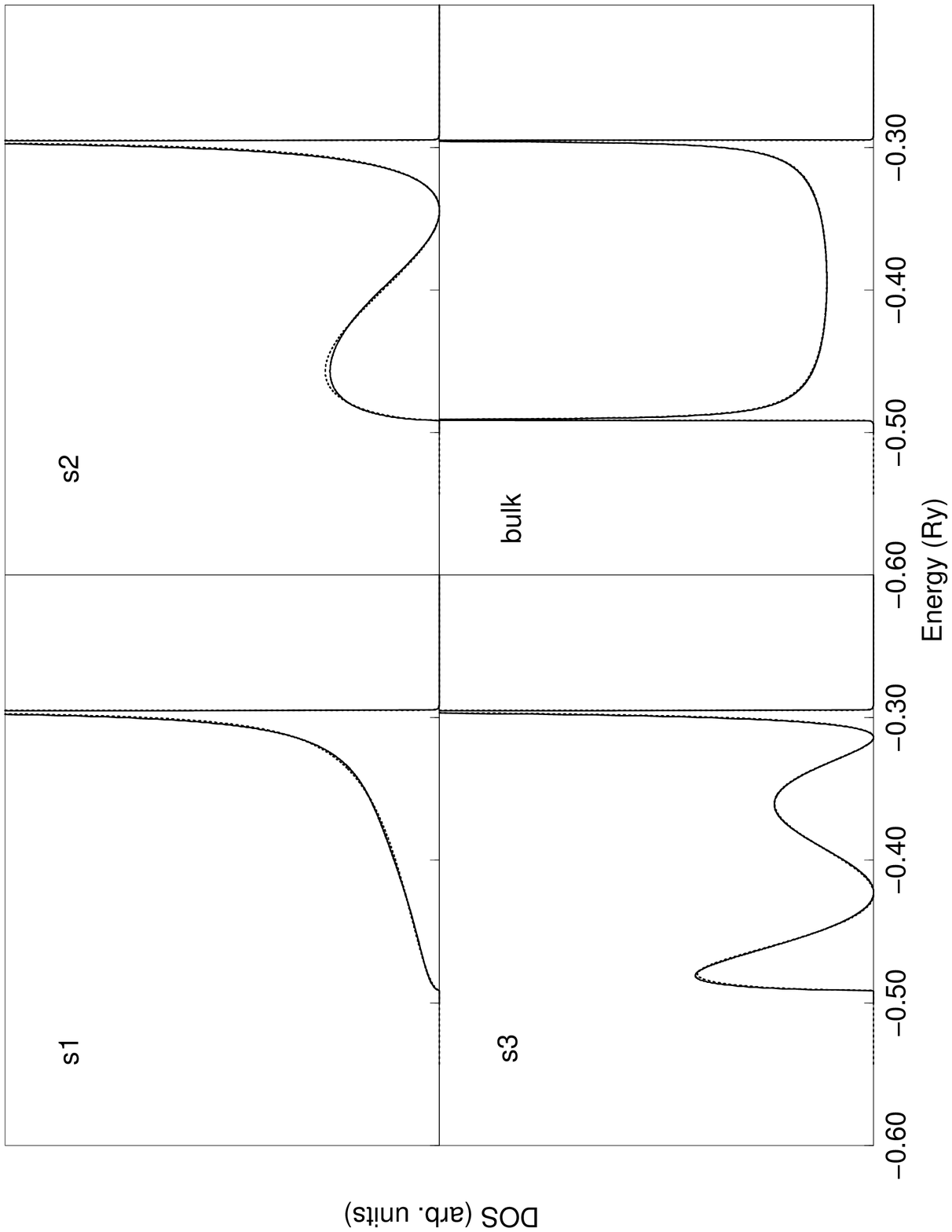}}
\end{picture}
\vspace*{-1.5cm}
\begin{figure}[h]
\caption{
The solid lines are the calculated
$xy$ contributions to the ${\bf k_{||}}$-, and layer-resolved DOS at
$\bf k_{||}=0$ for a Rh(001) surface. The topmost three sample layers
 are denoted
s1, s2, and  s3. The dotted lines are the local density
of states of the terminal, and first two interior neighbors of a perturbed
semiinfinite chain with nearest neighbor coupling $2t= 0.0980$ Ry, and
potential shifts
$V_1/2t= 0.4984$ and $V_2/2t=0.0010$    
on the terminal and penultimate sites. The position of
the bulk substrate Fermi level is at  $-0.0765$ Ry.}\label{fig3}
\end{figure}

\newpage
\vspace*{4.5cm}
\begin{picture}(3,10)
\put(.0,.0){\includegraphics{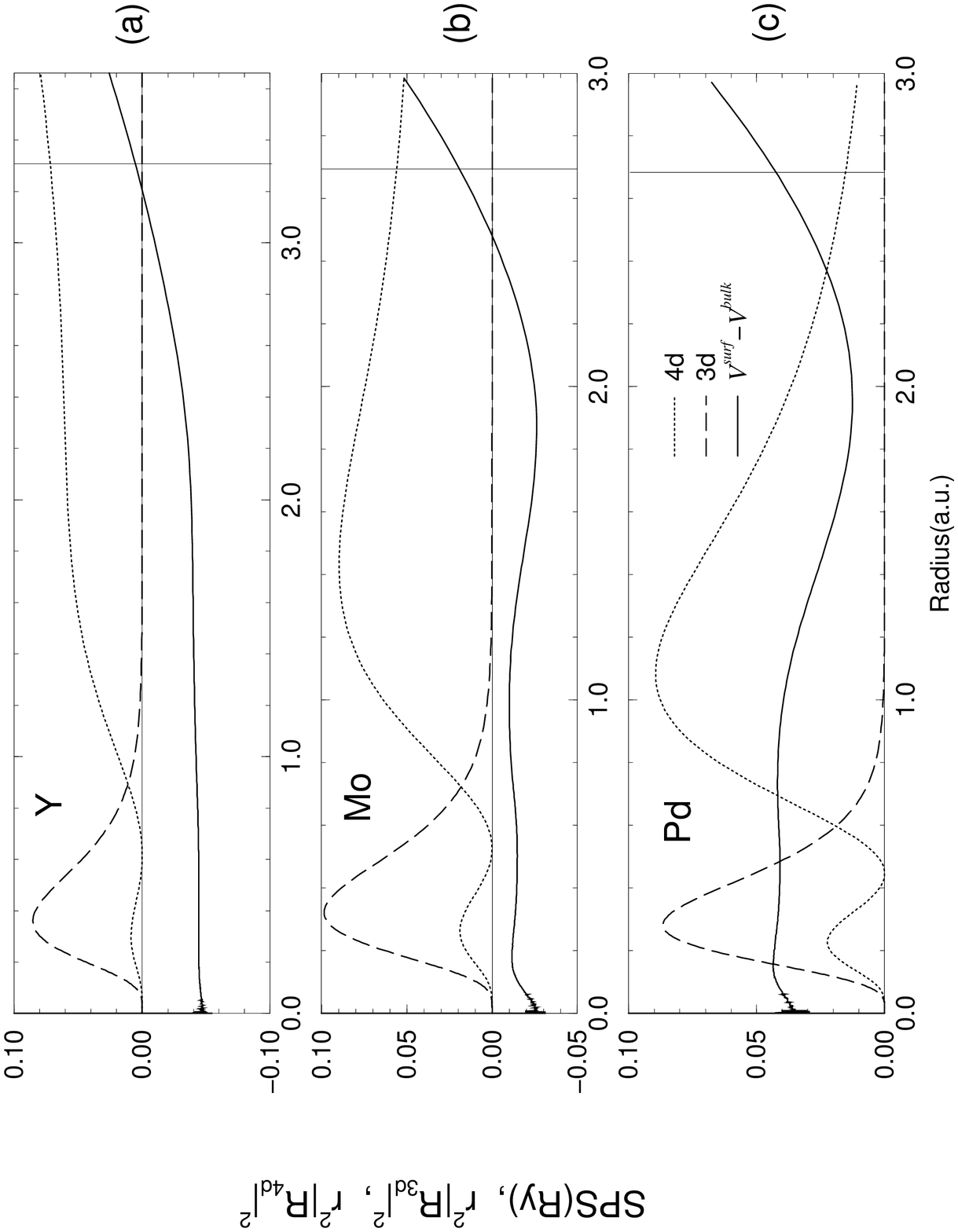}}
\end{picture}
\vspace*{-1.5cm}
\begin{figure}[h]
\caption{
Spherically symmetric part of the surface potential shift and 
the radial solutions of the Schr\"odinger
equation within the corresponding atomic-sphere in the bulk,
$r^2|R_{nl}(r)|^2$, $nl=3d, 4d$
for (a) Y, (b) Mo, and (c) Pd.
The vertical lines correspond to a radius of half the nearest neighbor separation.}\label{fig4}
\end{figure}

\newpage
\vspace*{4.5cm}
\begin{picture}(3,10)
\put(.0,.0){\includegraphics{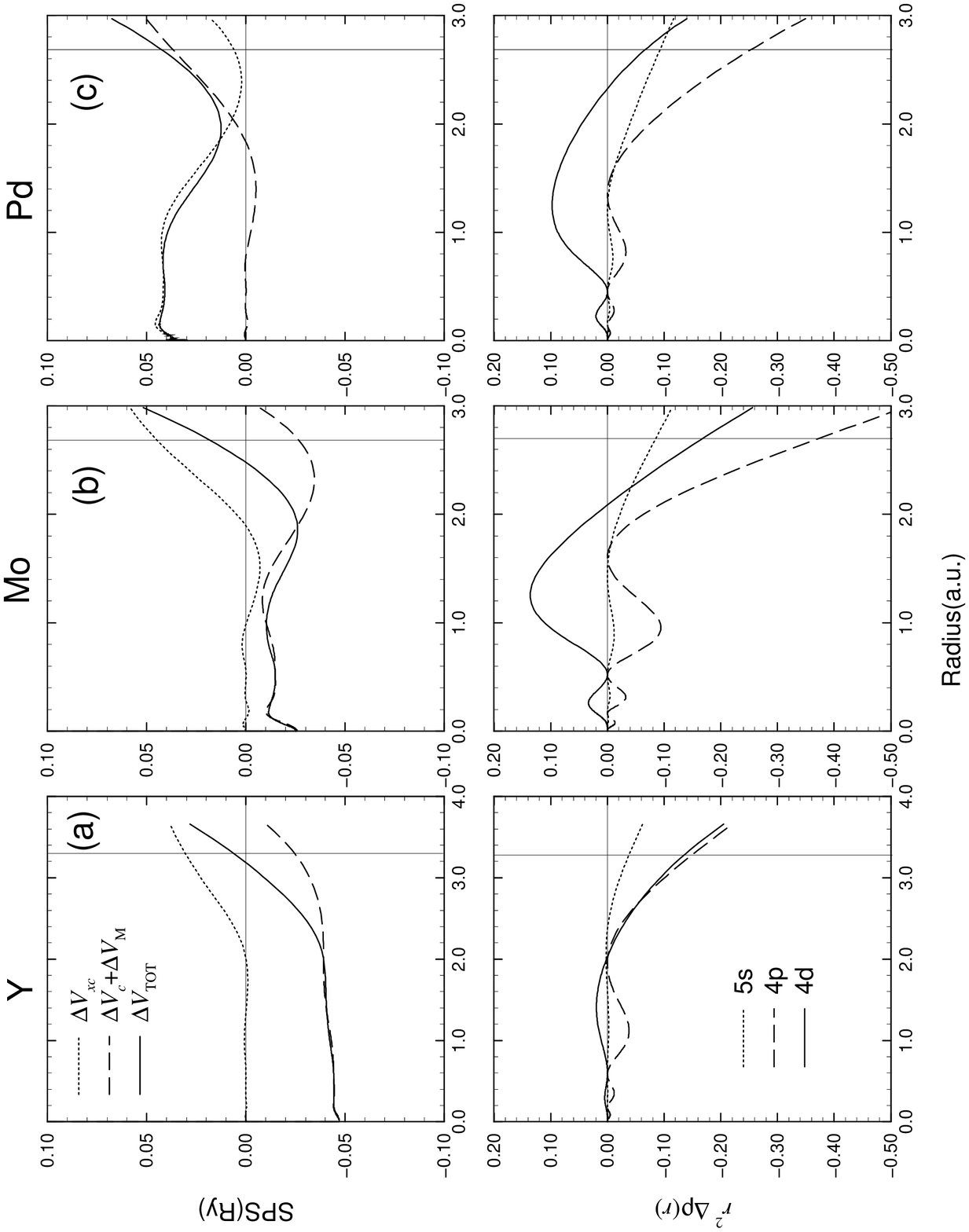}}
\end{picture}
\vspace*{-1.5cm}
\begin{figure}[h]
\caption{
Spherically symmetric part of the contributions to the surface potential shift and 
the 
$r^2 \Delta\rho_l(r)$, $l=s,p,d$
of the surface atoms for (a) Y, (b) Mo, and (c) Pd.
The vertical lines correspond to a radius of half the nearest neighbor
separation.}\label{fig5}
\end{figure}

\end{document}